\documentclass{aa}
\usepackage{graphicx}
\usepackage{txfonts}

\begin{document}

\title{On the importance of the wind emission to the \\optical continuum of OB supergiants}

\author{M. Kraus\inst{1} \and J. Kub\'at\inst{1}
    \and
J. Krti\v{c}ka\inst{2} 
  }

\institute{Astronomick\'y \'ustav, Akademie v\v{e}d \v{C}esk\'e republiky, Fri\v{c}ova 298, 251~65 Ond\v{r}ejov, Czech Republic\\
 \email{kraus@sunstel.asu.cas.cz; kubat@sunstel.asu.cas.cz}
 \and
  \'Ustav teoretick\'e fyziky a astrofyziky P\v{r}F MU, 611~37 Brno, Czech Republic\\
 \email{krticka@physics.muni.cz}
      }

\date{Received; accepted}


\abstract
{Thermal wind emission in the form of free-free and free-bound emission is
known to show up in the infrared and radio continuum of hot and massive stars.
For OB supergiants with moderate mass loss rates and a wind velocity 
distribution with $\beta\simeq 0.8\ldots 1.0$, no influence of the wind to
the optical continuum, i.e. for $\lambda \la 1.0\,\mu$m, is expected.  
Investigations of stellar and wind parameters of OB supergiants over the last
few years suggest, however, that for many objects $\beta$ is much higher than 
1.0, reaching values up to 3.5.}
{We investigate the influence of the free-free and free-bound emission on the 
emerging radiation, especially at optical wavelengths, from OB supergiants 
having wind velocity distributions with $\beta \ge 1.0$.}
{For the case of a spherically symmetric, isothermal wind in local 
thermodynamical equilibrium (LTE) we calculate 
the free-free and free-bound processes and the emerging wind and total continuum
spectra. We localize the generation region of the optical wind continuum and 
especially focus on the influence of a $\beta$-type wind velocity 
distribution with $\beta > 1$ on the formation of the wind continuum at
optical wavelengths.}
{The optical wind continuum is found to be generated within about $2\,R_{*}$
which is exactly the wind region where $\beta$ strongly influences the density 
distribution. We find that for $\beta > 1$, the continuum of a typical OB 
supergiant can indeed be contaminated with thermal wind emission, 
\protect{\it even at optical wavelengths}. The strong increase in the optical 
wind emission is dominantly produced by free-bound processes.}
{}



\maketitle

\section{Introduction}

It is well established that for massive stars the appearance of (thermal) 
excess emission at infrared (IR) and radio wavelengths is caused by free-free 
and (to a small fraction also by) free-bound emission generated in their winds 
(see e.g. Panagia \& Felli \cite{PanagiaFelli}; Olnon \cite{Olnon}). At 
radio wavelengths, the free-free excess emission is usually used to derive 
the mass loss rates of hot and massive stars (see e.g. Lamers \& Leitherer 
\cite{LamersLeitherer}; Puls et al. \cite{Puls96}). 

Waters \& Lamers (\cite{WatersLamers}) have investigated this excess emission 
in detail, especially in the near-IR region. These authors studied the 
influence of free-free and free-bound emission to the total continuum and 
pointed already to the importance of the wind velocity (and 
hence the wind density) distribution that can severely alter the wind 
contribution in the IR. 

However, the investigations of Waters \& Lamers (\cite{WatersLamers}) were
restricted to the IR and radio range, i.e. they calculated the wind contribution
for $\lambda \ga 1\,\mu$m only, while during earlier studies Brussaard \& van 
de Hulst (\cite{Brussaard}) had noted that free-bound processes might become 
very important in the optical and UV range for temperatures typically found
in the winds of hot stars and supergiants. 

Whether the free-bound emission in the wind indeed influences the optical 
continuum, depends severely on the density distribution. For line-driven winds,
the density follows from the equation of mass continuity, i.e. it is 
proportional to the mass loss rate, and inversely proportional to the wind
velocity. A high wind density can therefore be reached by either a high mass 
loss rate, or a low wind velocity. 

For massive stars with pronounced high mass 
loss rates like Wolf-Rayet stars, Luminous Blue Variables, or the group of 
B[e] stars, it is well known that the wind not only influences, but even 
dominates, the optical spectrum. In these objects, the wind is usually optically 
thick even in the visual range and thus completely hides the stellar spectrum. 
Some recent examples of the thermal wind influence in the form of free-free and
free-bound emission at optical wavelengths have been published, e.g. by Guo \& 
Li (\cite{Guo}) for the case of Luminous Blue Variables, and by Kraus et al. 
(\cite{Kraus07}) for a Magellanic Cloud B[e] supergiant.

The second density triggering parameter is the wind velocity distribution.
For OB-type stars with line-driven winds, the wind velocity is very often
approximated by a so-called $\beta$-law, where $\beta$ is in the range of 
$0.8\ldots 1.0$. Such a
$\beta$ value causes a rather fast wind acceleration at the base of 
the wind, and the wind reaches its terminal value within a few stellar radii
(see e.g. Lamers \& Cassinelli \cite{LamersCassinelli}). Consequently, the 
region of very high density that might cause enhanced free-bound emission is 
restricted to an extremely small volume around the stellar surface.
We can, therefore, expect that for OB-type stars, which have only moderate mass 
loss rates and whose wind velocity distributions have $\beta$ values between 
0.8 and 1.0, there will be no noticable influence of the wind on their optical 
continuum emission.

During the last few years, huge effort has been made to determine precisely
the stellar and wind parameters of OB-type supergiants in the Galaxy (e.g. 
Kudritzki et al. \cite{Kudritzki}; Markova et al.\,\cite{Markova}; Fullerton 
et al. \cite{Fullerton}; Crowther et al. \cite{Crowther}; Puls et al. 
\cite{Puls}), in the Magellanic Clouds (see e.g. Evans et al. \cite{Evans}; 
Trundle et al. \cite{Trundle}; Trundle \& Lennon \cite{TrundleLennon}), and 
beyond, e.g., in M\,31 (e.g. Bresolin et al. \cite{Bresolin}). Interestingly, 
many OB supergiants are found to have rather high $\beta$ values (up to $\beta 
= 3.5$, see Sect.\,\ref{highbeta}). Such high $\beta$-values, resulting in a 
strong density increase close to the stellar surface due to a much slower 
wind acceleration, should have noticable effects on the thermal wind emission 
not even in the near-IR, but extending also to optical wavelengths. In this 
paper, we, therefore, aim to investigate and discuss in detail the influence 
of high $\beta$ values on the wind emission of OB supergiants, especially at 
optical wavelengths.

\section{Description of the model OB supergiant}\label{ff}
                                                                                
The calculation of the continuum emission of a typical OB supergiant is
performed in three steps: (i) first we calculate the stellar emission of the
supergiant with no stellar wind, (ii) then, we calculate the emission of the
wind with the stellar parameters as boundary conditions, (iii) and finally,
we combine the two continuum sources whereby the stellar emission still has
to pass through the absorbing wind. A justification of the use of this
so-called core-halo approximation together with a discussion of several other
assumptions and simplifications are given in Sect.\,\ref{disc}.

\subsection{The stellar model}

To simulate a typical OB supergiant we adopt the following set of stellar
parameters: $T_{\rm eff} = 33\,000$\,K; $R_{*} = 17.2\,R_{\odot}$; $\log
L_{*}/L_{\odot} = 5.5$; and $\log g = 3.4$. With these parameters, we compute
the stellar continuum emission of a hydrogen plus helium atmosphere,
given by the Eddington flux, $H_{\nu}$. These calculations are performed 
with the code of Kub\'at (\cite{Kubat03}, and references therein), which is 
suitable for the calculation of non-LTE spherically-symmetric model 
atmospheres in hydrostatic and radiative equilibrium. 

\subsection{The wind model}\label{windmodel}

\begin{figure}[t!]
\resizebox{\hsize}{!}{\includegraphics{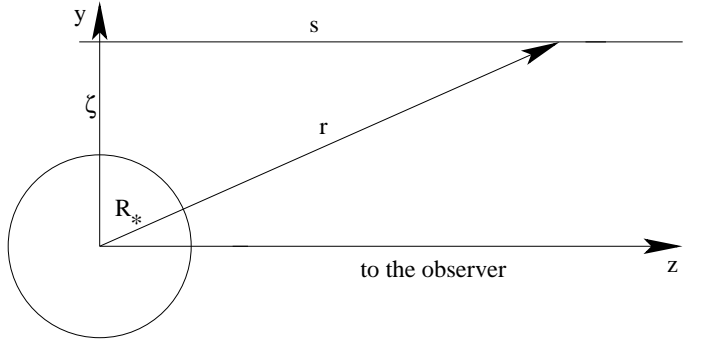}}
\caption{Frame of reference used for the calculations (after Panagia \& Felli
\cite{PanagiaFelli}). The $x$-axis is perpendicular to the drawing-plane.}
\label{sketch}
\end{figure}

For simplification, we assume that hydrogen in the wind is fully ionized 
while further contributions to the electron density distribution from, e.g., 
helium and the metals, are neglected. This means that for a given mass loss rate
the real number density of free electrons is underestimated. This leads to an 
underestimation of the total wind emission generated via free-free and 
free-bound processes. However, since we are interested only in the effect on 
the emerging wind emission caused by different velocity distributions, such a 
simplification is reasonable. We further neglect electron scattering, and 
describe the wind zone following Panagia \& Felli (\cite{PanagiaFelli}) with a 
spherically symmetric stationary model in LTE. We show the frame of reference 
used for our computations in Fig.\,\ref{sketch}. The electron number density 
distribution, $n_{\rm e}(r)$, then equals the hydrogen number density 
distribution, $n_{\rm H}(r)$, which follows from the equation of mass 
continuity,
\begin{equation}
n_{\rm e}(r) = n_{\rm H}(r) = \frac{\dot{M}}{4\pi\mu m_{\rm H}r^{2}\varv(r)}\,
.
\label{masscont}
\end{equation}
This equation relates the density at any location $r$ in the wind to the mass 
loss rate, $\dot{M}$, of the star and the wind velocity, $\varv(r)$. The 
parameters $\mu$ and $m_{\rm H}$ are the mean atomic weight, 
for which we use a value of 1.4 (i.e. solar composition), 
and the atomic hydrogen mass, 
respectively.

The velocity increase in line-driven winds of hot stars is approximated
with a $\beta$-law of the form (see e.g. Lamers \& Cassinelli
\cite{LamersCassinelli})
\begin{equation}
\varv(r) = \varv_{0} + (\varv_{\infty} - \varv_{0}) \left( 1 -
\frac{R_*}{r}\right)^{\beta}
\label{velo}
\end{equation}
where $R_*$ is the stellar radius, $\varv_{\infty}$ is wind terminal velocity,
and $\beta$ describes the 'steepness` of the velocity increase at the base of
the wind. The term $\varv_{0}$ gives the velocity at the base of the wind, i.e.,
at $r = R_{*}$, for which we use the thermal hydrogen velocity given by 
the wind temperature.

The wind temperature for hot stars is known to drop quickly with distance 
(see e.g., Drew \cite{Drew}), and a useful description of the radial electron
temperature distribution as a function of effective temperature and wind 
velocity has been derived by Bunn \& Drew (\cite{Bunn})
\begin{equation}\label{temp_dist}
T_{\rm e}(r) = 0.79\,T_{\rm eff} - 0.51\,\frac{\varv(r)}
{\varv_{\infty}}\,T_{\rm eff}\, .
\end{equation}
This relation states that in the vicinity of the stellar surface, the electron
temperature is roughly $0.8 T_{\rm eff}$, 
a value that has been found and confirmed by other investigators as well
(see e.g. Drew \cite{Drew}; de Koter \cite{deKoter}; Krti\v{c}ka 
\cite{Krticka}).

For the purpose of our investigation, it is sufficient to restrict the model
computations to an 
isothermal wind, and throughout this paper we will use the value of $T_{\rm e} 
= T_{\rm e}(R_*) = 25\,760$\,K (which is lower than $0.8 T_{\rm eff}$)
following from Eq.\,(\ref{temp_dist}) for our 
star with $T_{\rm eff} = 33\,000$\,K. A justification for these assumptions 
and a discussion of their reliability are given in Sect.\,\ref{tempjust}.

The assumption of a constant wind temperature allows for a simplified treatment 
of the radiation transfer and therefore of the intensity calculation. 
Consequently, at each impact parameter, $\zeta$ (see Fig.\,\ref{sketch}), the 
intensity of the wind emission is given by
\begin{equation}
I_{\nu} = B_{\nu}(T_{\rm e})\left( 1 - e^{-\tau_{\nu}(\zeta)}\right)\,.
\label{intensity}
\end{equation}
The optical depth is defined as the line-of-sight integral over the absorption 
coefficient of the free-free and free-bound processes, $\kappa_{\nu}(\zeta,s)$,
\begin{equation}
\tau_{\nu}(\zeta) = \int\limits_{s_{\rm min}}^{s_{\rm max}}\kappa_{\nu}
(\zeta,s)\,ds
\end{equation}
with the integration limits
\begin{eqnarray}
s_{\rm min} & = & \sqrt{R_{*}^{2} - \zeta^{2}} \qquad \qquad \textrm{for} \qquad
0 \le \zeta < R_{*}\, , \\ 
s_{\rm min} & = & -\sqrt{R_{\rm out}^{2} - \zeta^{2}} \qquad \textrm{for} 
\qquad R_{*} \le \zeta < R_{\rm out}\, ,
\end{eqnarray}
and
\begin{equation}
s_{\rm max} = \sqrt{R_{\rm out}^{2} - \zeta^{2}}\, ,
\end{equation}
where $R_{\rm out}$ is the outer edge of the ionized wind.
The total observable flux at earth of the wind continuum emission follows
from the integration of the wind specific radiation intensity (\ref{intensity})
over the wind zone projected to the sky
\begin{equation}
F_{\nu, {\rm wind}} = \frac{2\pi}{d^{2}}\int\limits_{0}^{\zeta_{\rm
max}} B_{\nu}(T)\left( 1 - e^{-\tau(\zeta)}\right) \zeta d\zeta
\label{flux}
\end{equation}
where $\zeta_{\rm max} = R_{\rm out}$, and $d$ is the distance to the object.

With the wind absorption along each line of sight, i.e., each impact parameter
$\zeta$, the stellar flux passing through the wind zone can be calculated from, 
\begin{equation}
F_{\nu, {\rm star}} = \frac{\pi R_{*}^{2}}{d^{2}}~I_{\nu}~e^{-\tau_{\nu}(\zeta 
= 0)}\, ,
\label{stelatt}
\end{equation}
with the stellar intensity $I_{\nu} = 4\,H_{\nu}$. For simplicity, we 
restrict the calculation of the wind attenuation to the direction
$\zeta = 0$, only. This delivers a lower limit to the real attenuation so
that we slightly overestimate the stellar flux leaving the wind zone.
 
Finally, the total continuum emission is the sum of attenuated stellar and 
wind contribution, i.e. 
\begin{equation}
F_{\nu} = F_{\nu, {\rm star}} + F_{\nu, {\rm wind}}\, .
\end{equation}

\subsubsection{Gaunt factors for free-free and free-bound processes}\label{gauntfactors}

The absorption coefficient of the free-free and free-bound processes,
$\kappa_{\nu}(\zeta,s)$ (in cm$^{-1}$), is given by (see e.g. Brussaard \& 
van de Hulst \cite{Brussaard})
\begin{equation}
\kappa_{\nu}(\zeta,s) \simeq 3.692\cdot 10^{8}\,\frac{n_{\rm e}
(\zeta,s)^{2}}{\nu^{3}T^{1/2}} \left(1-e^{-\frac{h\nu}{kT}}\right)\left(g_{\rm
ff}(\nu,T)+ f(\nu,T) \right)
\label{kappa_red}
\end{equation}
where $g_{\rm ff}(\nu,T)$ is the Gaunt factor for free-free emission and
the function $f(\nu,T)$ contains the Gaunt factors for free-bound processes.
To calculate the absorption coefficient, we need to specify the Gaunt factors 
for both the free-free and the free-bound processes.

In the literature there exist several approximations for the calculation of the free-free 
Gaunt factors in either the short or the long wavelength regime. To allow for 
an appropriate transition from one regime to the other (see e.g. Waters \& 
Lamers \cite{WatersLamers}; Kraus \cite{PhD}) we use in the long-wavelength 
region the relation of Allen (\cite{Allen}), and in the short-wavelength region 
the expression of Gronenschild \& Mewe (\cite{Gronenschild}). The resulting
Gaunt factors are in good agreement with those calculated from the 
approximation given by Mihalas (\cite{Mihalas}), which is based on calculations 
of Berger (\cite{Berger}).

\begin{figure}[t!]
\resizebox{\hsize}{!}{\includegraphics{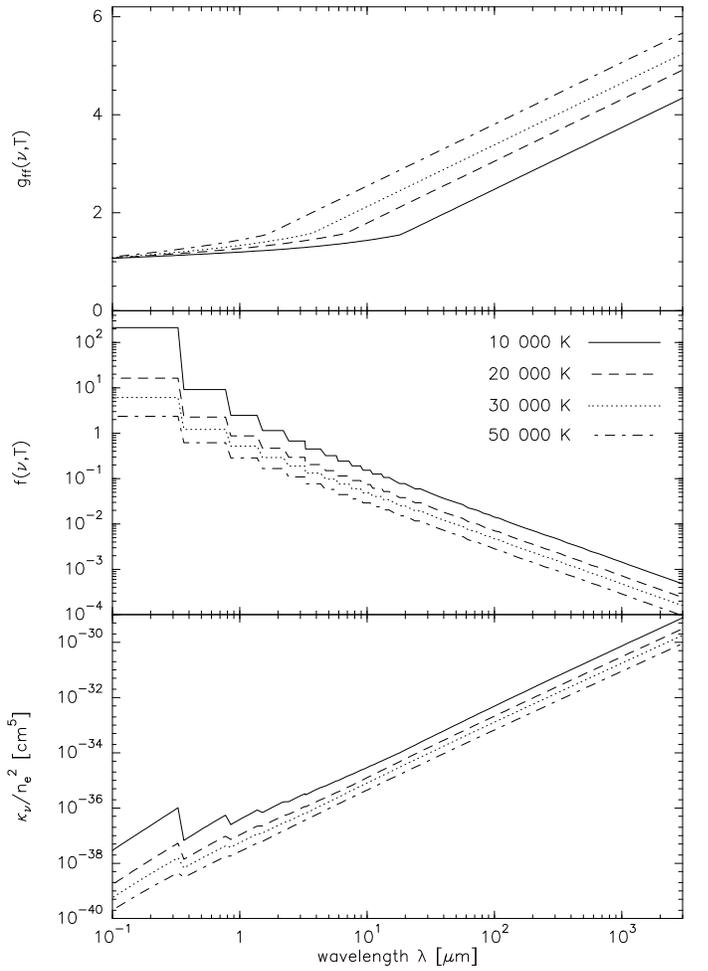}}
\caption{
{\em Top:}
Gaunt factors for free-free processes. They converge to unity in the optical,
but increase with wavelength in the IR and radio range.
{\em Middle:}
The function $f(\nu,T)$ containing the Gaunt factors for the free-bound 
processes. The values of $f(\nu,T)$ increase steeply in the optical, while for 
$\lambda > 1\,\mu$m they drop quickly and become unimportant compared to the 
free-free Gaunt factors.
{\em Bottom:}
Absorption coefficient of the free-free and 
free-bound processes (see Eq.\,(\ref{kappa_red})). Clearly visible is the 
growing influence of the free-bound processes in the optical range, especially 
for decreasing temperature. The line styles indicate the electron temperature.} 
\label{gaunt}
\end{figure}

In the top panel of Fig.\,\ref{gaunt}, we plotted the free-free Gaunt factors 
calculated over a large range of wavelengths and for different electron 
temperatures. The curves converge to unity for wavelengths shorter than $\sim 
1\,\mu$m, while they increase with wavelength in the IR and radio range, where 
free-free processes are known to dominate the continuum emission. The free-free 
Gaunt factors thereby depend only weakly on the chosen electron temperature.

The situation is completely different for the  
function $f(\nu,T)$, which itself is not a Gaunt factor but that contains the
Gaunt factors of the free-bound processes. We calculate this function with
Eq.\,(25) of Brussaard \& van de Hulst (\cite{Brussaard}). The factors $g_{n}$
entering this equation are thereby the Gaunt factors for the transitions from 
the free level $E = h\nu_{0}/n^2 + h\nu$ to the bound level $E = h\nu_{0}/n^2$.
These individual Gaunt factors can, in principle, be calculated with the exact
formula given by Menzel \& Pekeris (\cite{Menzel}) or by the approximation 
provided by Mihalas (\cite{Mihalas}). But it turns out, that these Gaunt 
factors have values between 0.8 and 1.1 in the wavelength range 
of our interest (i.e. for $\lambda \ga 0.1\,\mu$m) and converge to unity 
for $\lambda \ga 10\,\mu$m (see Fig.\,8 of Brussaard \& van de Hulst 
\cite{Brussaard}). It is therefore reasonable to use $g_{n}
\simeq 1$ for all levels of interest. With this approximation, we calculated
the function $f(\nu,T)$ over the same wavelength range and for the same 
electron temperatures as the free-free Gaunt factors. The results are shown in 
the middle panel of Fig.\,\ref{gaunt}.

For $\lambda \ga 1\,\mu$m, the values of $f(\nu,T)$ are $\ll 1$. The hydrogen
free-bound processes, therefore, play no role in the IR and radio regimes.
At optical wavelengths however, i.e., for $\lambda < 1\,\mu$m, $f(\nu,T)$
starts to grow steeply, with $f(\nu,T) \gg 1$ especially for electron 
temperatures below $\sim 30\,000$\,K, i.e., for values typically found in
the winds of OB supergiants. This effect has been mentioned already
by Brussaard \& van de Hulst (\cite{Brussaard}) and Kraus (\cite{PhD}) who
pointed to the possible importance of the contribution of free-bound processes
to the total continuum emission.

The resulting absorption coefficient, $\kappa_{\nu}$, is shown in the bottom
panel of Fig.\,\ref{gaunt} where we plot $\kappa_{\nu}/n_{\rm e}^2$ 
calculated with Eq.\,(\ref{kappa_red}) for the different electron temperatures. 
Besides the well-known increase with increasing wavelength, the absorption 
coefficient also peaks at short wavelengths due to the growing influence of the 
free-bound processes.

\section{Results}

We fix the terminal wind velocity at ${\varv}_{\infty} = 1400$\,km\,s$^{-1}$ 
and the mass loss rate at $\dot{M} = 5\times 10^{-6}\,M_{\odot}$yr$^{-1}$. 
In addition, we place the object to an arbitrary distance of 1\,kpc. 
The resulting continuum emission is first calculated for the case of
$\beta = 1.0$ (Sect.\,\ref{beta_1}) before we turn to the more
interesting case of higher beta values in Sect.\,\ref{highbeta}.

\subsection{A wind with $\beta = 1.0$}\label{beta_1}

\begin{figure}[t!]
\resizebox{\hsize}{!}{\includegraphics{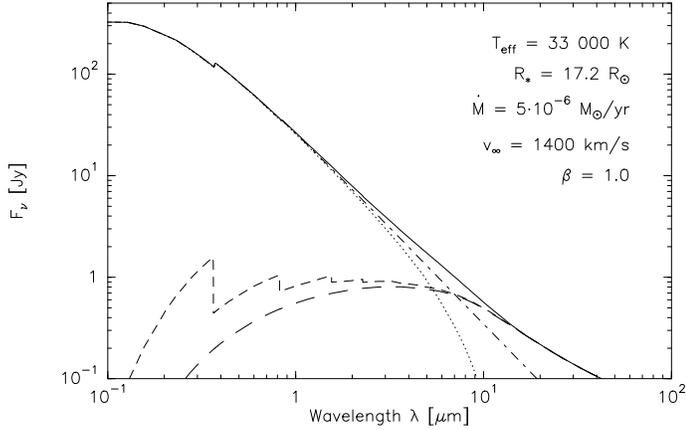}}
\caption{Continuum emission of a typical OB supergiant (solid line), consisting
of the stellar atmosphere having passed through the absorbing wind (dotted)
and the thermal wind emission (dashed). 
Also included is the emission from a wind with pure free-free processes 
(long-dashed line) and the  
emission of the star without wind (dashed-dotted).}
\label{beta1}
\end{figure}

\begin{figure}[t!]
\resizebox{\hsize}{!}{\includegraphics{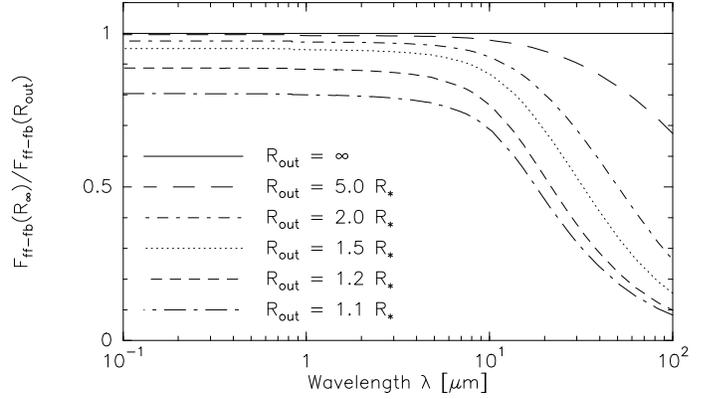}}
\caption{Increase of the fluxes generated within wind zones with outer edge
$R_{\rm out}$, with respect to an infinitely large wind zone. The thermal wind
emission in the optical and near-IR is generated completely in the vicinity of
the stellar surface, i.e., within $5\,R_{*}$ only.}
\label{generation}
\end{figure}

For many hot star winds, $\beta$ lies typically in the range $0.8\ldots 1.0$
(e.g. Markova et al.\,\cite{Markova}; Repolust et al.\,\cite{Repolust}), and we
use $\beta = 1.0$ for our first test calculation, which serves as a reference
model. The resulting spectrum is shown in Fig.\,\ref{beta1} where we plot the
total continuum emission of our test OB supergiant, consisting of the
stellar atmosphere having passed through the absorbing wind and the thermal 
wind emission. For comparison, we also included the emission of the star
without wind. It is obvious, that the wind
influences the total spectrum only for $\lambda \ga 1\,\mu$m, while the optical
spectrum remains uninfluenced. The stellar continuum only suffers in the IR and
radio range, where for $\lambda \ga 10\,\mu$m the stellar emission is
completely absorbed by the wind.
In Fig.\,\ref{beta1}, we also included the results for a wind with pure 
free-free emission. From a comparison of the 
total wind emission with the pure free-free wind emission, it is obvious that 
free-bound processes dominate the wind emission in the optical and near-IR 
wavelengths, i.e., for $\lambda\la 2\,\mu$m.   

To understand where the optical wind continuum is generated, we re-calculate 
the wind emission for different values of the outer edge of the wind zone. We 
then compared the resulting emission to the emission from an infinitely large 
wind, and we plotted the ratios versus frequency in Fig.\,\ref{generation}. 
From this plot, it is evident that the wind emission in the optical and near-IR 
is generated in the vicinity of the stellar surface, i.e., within $5\,R_{*}$. 
Most of the emission for $\lambda\lesssim10\,\mu$m (about 95\,\%) is even 
generated within the innermost $1.5\,R_{*}$.

\subsection{Winds with $\beta > 1.0$}\label{highbeta}

In recent years, detailed investigations of the wind parameters of OB
supergiants have revealed that for many objects the $\beta$ parameter in the 
velocity law given by Eq.\,(\ref{velo}) varies over a much larger range. Even 
values as high as 3.5 are reported. A list of $\beta$ values of OB supergiants 
found in the literature is provided in Table\,\ref{beta_values}.

\begin{table}[t!]
\caption{Literature $\beta$ values for OB supergiants. References are:
M04 = Markova et al. (\cite{Markova}); C06 = Crowther et al. (\cite{Crowther});
K99 = Kudritzki et al. (\cite{Kudritzki}); E04 = Evans et al. (\cite{Evans});
T05 = Trundle \& Lennon (\cite{TrundleLennon}); T04 = Trundle et al. 
(\cite{Trundle}).}
\label{beta_values}
  \begin{tabular}{lccc}
    \hline
    \hline
  Galaxy & Sp.Types & $\beta$ & Ref. \\  
   \hline
  Milky Way & O4 -- O9.7 & 0.7 -- 1.25 & M04 \\
  Milky Way & O9.5 -- B3 & 1.2 -- 3.0 & C06 \\
  Milky Way & B0 -- B3 & 1.0 -- 3.0 & K99 \\
  Magellanic Clouds & O8.5 -- B0.5 & 1.0 - 3.5 & E04 \\
  Magellanic Clouds & B0.5 -- B2.5 & 1.0 - 3.0 & T05 \\
  Magellanic Clouds & B0.5 -- B5 & 1.0 - 3.0 & T04 \\
   \hline
  \end{tabular}
\end{table}

\begin{figure}[t!]
\resizebox{\hsize}{!}{\includegraphics{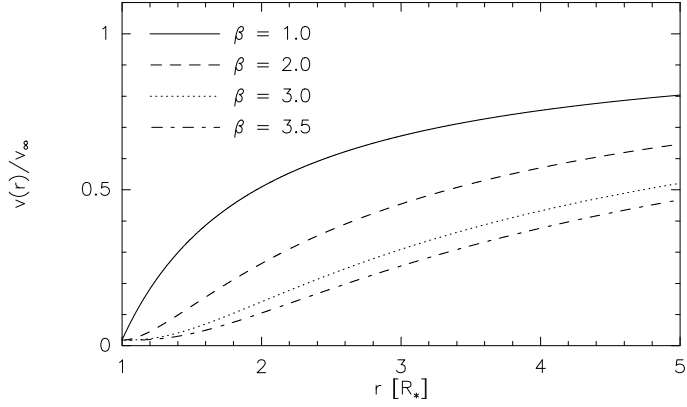}}
\caption{Velocity increase in winds with different $\beta$ values. The higher
the $\beta$, the more slowly the wind is accelerated and the farther away from 
the star it reaches its terminal value.}
\label{velo_beta}
\end{figure}

\begin{figure}[t!]
\resizebox{\hsize}{!}{\includegraphics{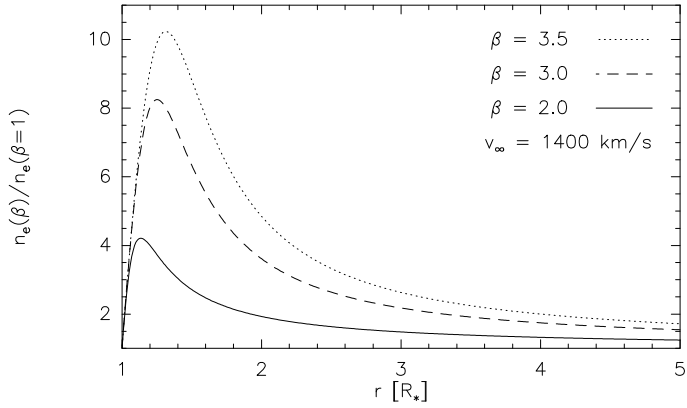}}
\caption{Density ratios close to the stellar surface for winds with $\beta > 
1.0$ with respect to the wind with $\beta = 1.0$. Due to the slower wind
acceleration for higher $\beta$, the wind density remains much higher over
a larger wind volume, resulting in a relative density enhancement close to
the stellar surface, i.e., within $2\ldots 3\,R_{*}$.}
\label{dens}
\end{figure}

The main effect of a higher $\beta$ value is the less steep increase in the
wind velocity with distance from the stellar surface. This is shown in 
Fig.\,\ref{velo_beta} where we plotted the velocity increase within $5\,R_{*}$ 
from the stellar surface in terms of the terminal velocity for different values 
of $\beta$. According to the equation of mass continuity 
(Eq.\,(\ref{masscont})), the wind density is proportional to $\varv(r)^{-1}$. 
Winds with higher $\beta$ values consequently have a (much) higher density in 
the accelerating wind regions.

We calculated the density distributions in winds with increasing $\beta$ and 
plotted the densities in terms of the density distribution for the 
wind with $\beta = 1.0$ in Fig.\,\ref{dens}. Close to the stellar surface
the densities in the winds with $\beta > 1$ are (much) higher than the 
reference value provided by the wind with $\beta = 1$. The higher the $\beta$, 
the stronger become these relative density enhancements due
to (much) slower wind acceleration (see Fig.\,\ref{velo_beta}). 
These relative density enhancements extend over the innermost $\sim 2\ldots
3\,R_{*}$ of the wind, while for larger distances, where the velocity 
distributions for the higher $\beta$ values also start to approach the terminal 
velocity, the densities converge. 

\begin{figure}[t!]
\resizebox{\hsize}{!}{\includegraphics{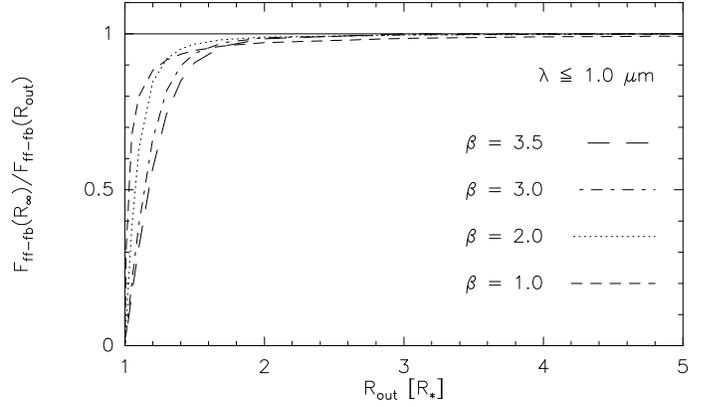}}
\caption{
Flux ratio $F_{\rm ff-fb}(R_\infty)/F_{\rm ff-fb}(R_{\rm out})$
as a function of outer wind edge $R_{\rm out}$ for different
values of $\beta$. The calculations shown are for $\lambda = 1\,\mu$m and are 
also valid for all smaller wavelengths. 
As reference, we include the ratio for $R_{\rm out} = R_{\infty}$ (solid
line).
For all values of $\beta$, the optical
wind continuum emission is generated well within the plotted wind size of 
$5\,R_{*}$.}
\label{ratio1mu}
\end{figure}

In Sect.\,\ref{beta_1} we showed that for a wind with $\beta = 1.0$ the optical 
continuum is generated within $< 2\,R_{*}$. This is exactly the range where 
$\beta$ has its strongest influence on the wind density. If the optical 
continuum for winds with $\beta > 1.0$ is generated within approximately
the same region, we can expect that the resulting free-bound emission will 
be strongly influenced by the value of $\beta$. 

\begin{figure}[t!]
\resizebox{\hsize}{!}{\includegraphics{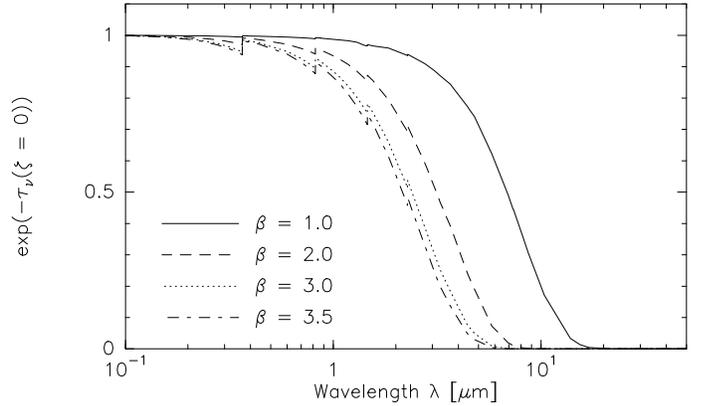}}
\caption{Fraction of stellar radiation escaping from the wind zone for different
values of $\beta$. While for $\beta = 1.0$ all the wind continuum emission for
$\lambda < 1\,\mu$m passes unabsorbed through the wind, the situation is
different for higher $\beta$ values for which even the optical wind continuum
suffers from wind absorption along its way through the wind. The higher $\beta$
the more stellar emission is absorbed also from the near- and mid-IR stellar
spectrum.}
\label{kappa}
\end{figure}

We determined the wind volume in which the optical flux is generated 
by comparing the flux generated within an infinitely large wind zone
with the flux generated within winds with different outer edges, $R_{\rm out}$.
The results for winds with different values of $\beta$ are shown as a function
of the outer edge, $R_{\rm out}$,
in Fig.\,\ref{ratio1mu}. These results are evaluated for $\lambda = 1\,\mu$m 
and are also valid for all smaller wavelengths (see Fig.\,\ref{generation}). 
Obviously, for winds with $\beta > 1.0$, more than 90\% of the optical 
continuum is generated within $1.5\,R_{*}$, and more than 98\% are produced 
within $2\,R_{*}$. 

Before we turn to the calculation of the total continuum emission from winds 
with $\beta > 1.0$, we first investigate the influence of $\beta$ on the
stellar emission. Since the wind absorption coefficient, $\kappa_{\rm ff-fb}$, 
is proportional to $n_{\rm e}^{2}$ (see Eq.\,(\ref{kappa_red})), an increasing 
density leads to a (strongly) increasing optical depth. The stellar flux will 
therefore be more strongly absorbed in the wind with a high $\beta$ compared to 
a wind with $\beta = 1$. This is illustrated in Fig.\,\ref{kappa}. Of course, 
the strongest depression of the stellar emission occurs at IR wavelengths 
($\lambda > 1\,\mu$m); with increasing $\beta$ also some part of the stellar 
flux at optical wavelengths, especially in the red part of the spectrum, is 
also absorbed.

\begin{figure}[t!]
\resizebox{\hsize}{!}{\includegraphics{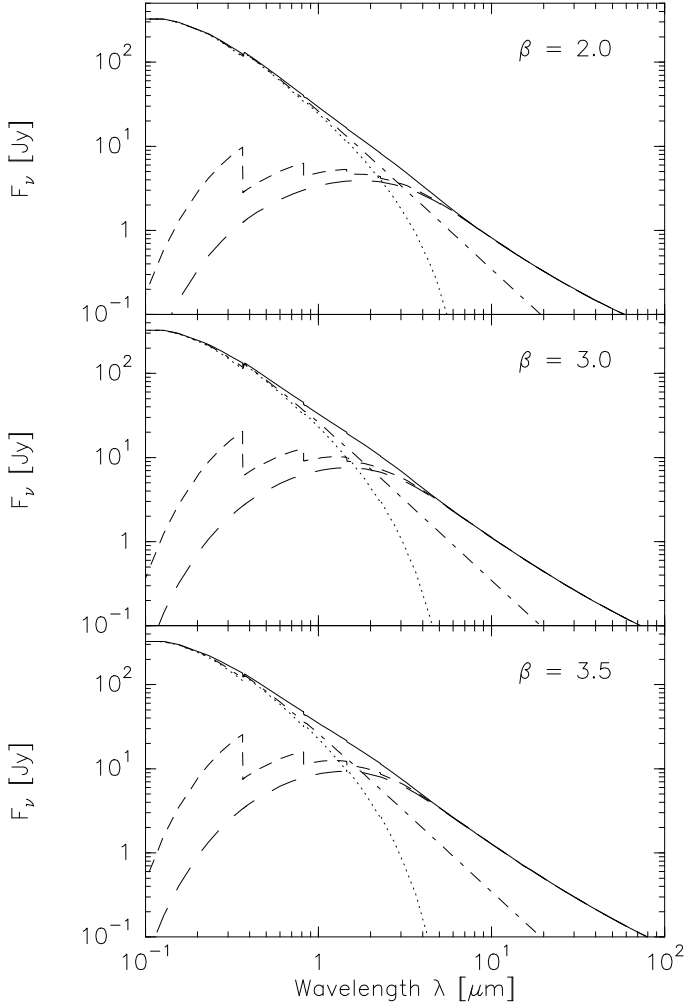}}
\caption{Comparison of total continuum spectra of the star plus wind system
(solid lines) for different values of $\beta$. 
As in Fig.\,\ref{beta1}, a wind with pure free-free emission is included in 
each panel (long-dashed line) to emphasize the growing importance of the 
free-bound contributions for $\lambda\le 2\,\mu$m.
With increasing $\beta$ the 
free-free and free-bound contributions (dashed) increase especially in the 
near-IR and optical spectrum. At the same time, the stellar continuum (dotted)
suffers from the increasing absorptivity of the wind zone. For comparison, the
pure stellar continuum is included (dashed-dotted).}
\label{beta}
\end{figure}

\begin{figure}[t!]
\resizebox{\hsize}{!}{\includegraphics{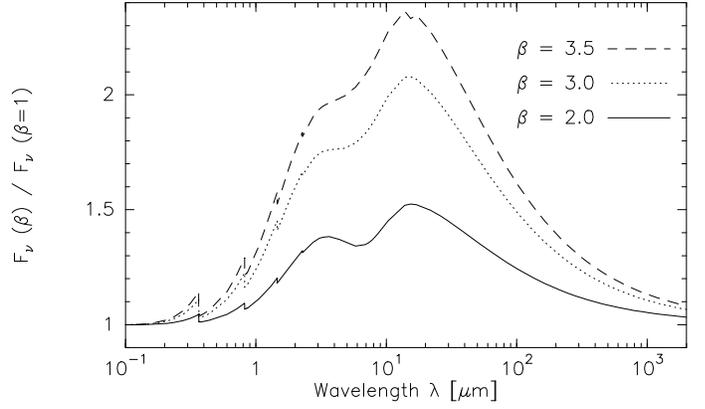}}
\caption{Continuum flux ratio of the star plus wind system for different
values of $\beta$ with respect to $\beta = 1.0$. With increasing $\beta$ the
increase in optical flux due to the increasing importance of the free-bound
emission from the wind becomes visible.}
\label{fluxratio}
\end{figure}

But while the increasing absorption coefficient results in a decrease 
of the stellar flux, it leads at the same time to a strong increase of the 
free-free and free-bound emission in the optical and IR part of the spectrum.
This can be seen upon inspection of Fig.\,\ref{beta} where we plotted the total
continuum emission for winds with different $\beta$ values. The optical wind
continuum increases with $\beta$ leading to an enhanced total continuum in
the IR and even at optical wavelengths. This increase, especially at optical
wavelengths, is more evident in Fig.\,\ref{fluxratio}. There we plot the flux 
ratio between the total continuum emission of the star plus wind system with 
$\beta > 1$ and the total continuum emission of the system with $\beta = 1$ as 
a function of wavelength. For $\beta\ga 3.0$, the total continuum
exceeds the one for $\beta = 1$ in the optical by about 15\% in the blue
spectral range and by about 25--30\% in the red spectral range, while in the
near- and mid-IR it reaches even more than 200\%.

\begin{figure}[t!]
\resizebox{\hsize}{!}{\includegraphics{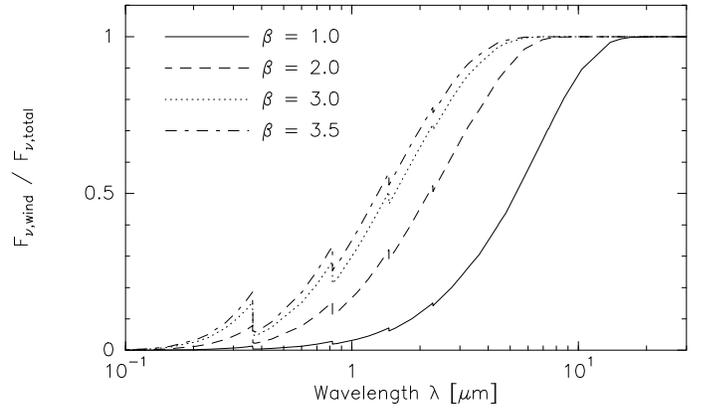}}
\caption{Ratio of the wind emission with respect to the total continuum
emission for different values of $\beta$. While for $\beta = 1.0$ the wind
contribution to the total optical continuum is less than 1\%, its influence
grows quickly with increasing $\beta$, reaching values up to 30\% in the
red part of the optical spectrum for $\beta > 3.0$.}
\label{windratio}
\end{figure}

The contribution of the pure wind emission to the total continuum production is 
finally shown in Fig.\,\ref{windratio}. Its importance grows and shifts to 
lower wavelengths with increasing $\beta$. While for $\beta = 1.0$ the 
contribution to the optical is less than 1\%, it grows to $> 10\,\%$ for $\beta 
= 2.0$ and to values exceeding even 30\% for $\beta \ga 3.0$. 

In summary, we can therefore state that {\it even at optical wavelengths} the 
wind plays a non-negligible role for the continuum emission of OB 
supergiants having wind velocity distributions with $\beta > 1.0$.

\section{Discussion}\label{disc}

For our calculations we made use of several assumptions and
simplifications like a constant wind temperature, the core-halo 
approximation, LTE even for the bound-free processes, and the neglect
of electron scattering. These are severe restrictions to keep the
model as simple as possible. Here, we want to discuss the influence and
possible consequences of these assumptions and simplifications on the model 
results. In addition, we want to address shortly the topic of wind 
clumping and its expected influence on our results.

\subsection{The wind temperature}\label{tempjust}

For our model calculations of stellar wind emission (see Sect.\,\ref{windmodel})
we made two severe assumptions concerning the temperature: (i) we assumed the 
wind to be isothermal, and (ii) we used a rather high temperature, i.e. the 
temperature at $r = R_*$, as the global electron temperature in the wind.
Here, we want to discuss why these two assumptions are reasonable.

Our choice of the electron temperature at $r = R_*$ means that we are
calculating a lower limit of the wind absorption coefficient, $\kappa_{\nu}$, 
because it is proportional to $T^{-1/2}$ and the Gaunt factors increase with 
decreasing temperature (see Fig.\,\ref{gaunt}). Consequently, according to the 
Eqs.\,(\ref{intensity}) and (\ref{flux}), the emission caused by free-free and 
free-bound processes, equally increases with decreasing electron temperature. 
Our results calculated for the high electron temperature thus underestimate the 
real wind continuum emission. The error in wind emission can be estimated
by looking at the real expected temperature distribution within the 
region where the optical wind continuum is generated.

\begin{figure}[t!]
\resizebox{\hsize}{!}{\includegraphics{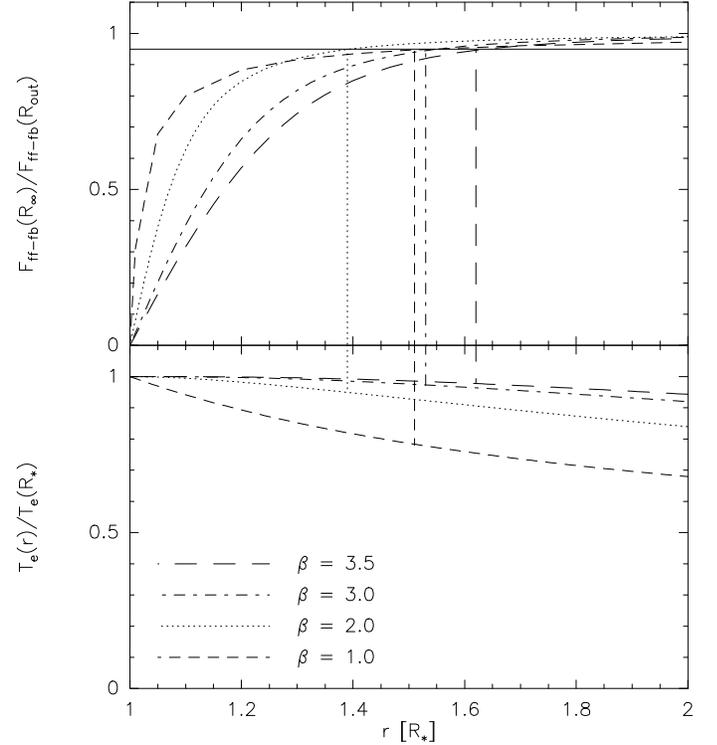}}
\caption{
{\em Top:}
As Fig.\,\ref{ratio1mu} but for the innermost $2\,R_{*}$, only. 
The solid horizontal line indicates the 95\,\% level, and the vertical lines 
the distances where, for winds of different $\beta$ values, this level is 
reached.
{\em Bottom:}
Electron temperature distribution within $2\,R_{*}$ for 
different values of $\beta$, normalized to the maximum electron temperature.
For $\beta > 1.0$, the inner parts of the winds remain much hotter 
and almost at a constant value compared to the wind with $\beta = 1.0$.}
\label{tempvert}
\end{figure}

During our investigations we found that for each value of $\beta$, 95\,\% of 
the wind continuum at optical wavelengths is generated in the vicinity of the 
stellar surface, i.e. within $1.40\ldots 1.65\,R_*$. This is shown in the 
top panel of Fig.\,\ref{tempvert}. The solid horizontal line in this plot 
indicates the 95\,\% level of the emission, and the vertical lines indicate, 
for each $\beta$, the distance at which this level is reached. 

To see how,in a more realistic wind model, the temperature would have changed
over the wind region in which the optical continuum emission is generated, 
we finally calculate the temperature distribution according to 
Eq.\,(\ref{temp_dist}), with the velocity distribution as defined by 
Eq.\,(\ref{velo}). The resulting temperature distributions for different values 
of $\beta$ are shown in the bottom panel of Fig.\,\ref{tempvert}. An increase 
in $\beta$ results in a lower wind velocity at the same distance from the star 
and, after Eq.\,(\ref{temp_dist}), in a higher wind temperature. Therefore, the 
higher the $\beta$, the hotter the wind remains at the same location. 

For winds with $\beta \ge 2.0$, the drop in wind temperature is found to be less 
than 5\,\% within the free-free and free-bound emission generation zone. Such a 
small decrease in temperature results in only a tiny and therefore negligible 
increase in wind emission. The assumption of an isothermal wind with $T_{\rm e} 
= T_{\rm e}(R_{*})$ is therefore well justified.

The situation is different for winds with $\beta = 1.0$. Here, the drop in wind 
temperature is found to be on the order of 20\,\%, which corresponds to a 
decrease in electron temperature by about 5\,000\,K for our chosen model star. 
In such a case, the temperature distribution in the wind is not negligible. The 
decrease in electron temperature results in an increase of the wind absorption 
coefficient of about a factor of two (see Fig.\,\ref{gaunt}), and consequently 
to a noticeably enhanced wind emission. However, even with an enhanced wind 
emission, the stellar spectrum still clearly dominates the total optical 
continuum. For OB supergiants with $\beta = 1.0$, the assumption of an 
isothermal wind with $T_{\rm e} = T_{\rm e}(R_{*})$ is therefore still
an acceptable approximation, as long as the mass loss rate of the star is not 
extremely high, as in the case of Luminous Blue Variables or B[e] 
supergiants.

\subsection{Core-halo approximation}

The construction of our star plus wind model is often referred to as the 
core-halo approximation, i.e., it is assumed that the stellar atmosphere is in 
hydrostatic equilibrium, and the wind is treated seperately with a density 
distribution following from the mass continuity equation. With such a treatment 
the transition between the atmosphere and the wind is not properly accounted 
for.

However, the intention of our research was to investigate the influence of one 
single parameter, namely the steepness of the velocity increase, $\beta$, 
keeping the stellar and the remaining wind parameters fixed. Deviations
in density distribution of the star plus wind system introduced by the use of
the core-halo approximation will consequently appear in all our discussed 
models and will influence all our results in the same way. But they will not 
significantly alter our conclusions, which are mainly based on the comparison 
of the spectra (i.e. the ratios) for stars with different $\beta$ values.

\subsection{Non-LTE effects and electron scattering}

In our calculations, we assumed the wind to be in LTE. This is a justified 
assumption for the free-free processes since they are collisional. 
For the bound-free processes the situation is less clear.
Here, non-LTE effects might play an important role.
According to our non-LTE model atmosphere calculations (e.g. Kub\'at
\cite{Kubat03}), departure coefficients can be (much) larger than unity
in the outer parts of the stellar atmosphere (especially for the ground
level) and, consequently, they lead to a higher bound-free opacity. 

We tested the influence of non-LTE effects on our results in a qualitative way 
by increasing artificially the bound-free opacity. This results in an increase 
in wind optical depth and, according to Eq.\,(\ref{stelatt}), to an increased 
attenuation of the stellar emission, while the wind emission remains largely 
unaffected. Consequently, the flux of the total continuum emission decreases 
with increasing bound-free opacity, i.e., increasing influence of non-LTE 
effects. At the same time, the importance of the wind emission with respect to 
the total continuum {\it increases}. The neglect of non-LTE effects thus 
results in a {\it lower limit} of the wind contribution. 

A similar conclusion can be drawn from the neglect of electron scattering.
Since electron scattering attenuates the stellar light passing through the
wind zone, it equally acts as an additional opacity source. The consequences
for the total continuum emission are therefore the same: a decrease in 
total emission (over the wavelength ranges that are dominated by the stellar
emission), and simultaneously an increase of the wind emission with respect
to the total continuum. 

For our model star plus wind system, we checked the electron scattering optical
depth in radial direction. We find that for the chosen stellar and wind 
parameters and the different $\beta$ values, the electron scattering optical
depth is always smaller than unity. However, compared to the free-bound
opacity, it is not negligible. This means, that we calculated indeed upper 
limits for the stellar emission resulting in lower limits for the wind 
emission with respect to the total continuum. A proper treatment of non-LTE 
effects and electron scattering can therefore only confirm our results: the 
importance of the wind contribution to the total optical continuum emission.

\subsection{Wind clumping}

In recent years, observational and theoretical investigations provided 
evidence of wind clumping (see e.g. Hillier \cite{Hillier05}). One of the most 
striking results found from detailed spectroscopic analyses (e.g. Crowther et 
al. \cite{Crowther02}; Hillier et al. \cite{Hillier}; Bouret et al. 
\cite{Bouret03,Bouret05}) was that if a wind is clumped, the mass loss rates 
inferred from spectroscopy might be lower on average by a factor of 3. In 
addition, Hillier et al. (\cite{Hillier}) found that clumping starts close to 
the photosphere, i.e., within the region in which the influence of $\beta$ is 
strongest.

The wind emission calculated in our study is not only determined by the $\beta$ 
parameter of the velocity distribution, but also by the mass loss rate of the
star that was kept constant during our analysis. Many of the authors who 
derived the high $\beta$ values (see Table\,\ref{beta_values}) also claim that 
their mass loss rates are derived under the assumption of unclumped winds and 
might well be a factor of 3 lower. Therefore, the question arises how our 
results might change if we account for wind clumping.
 
This paper is not aimed to study wind clumping in detail, instead we refer to 
the recent paper by Kraus et al. (\cite{Kraus08}) who investigated in more
detail the influence of wind clumping on the optical continuum emission of OB 
supergiants. Their results can be summarized as follows: wind clumping, 
introduced into the calculations, e.g., by the filling factor approach
of Hiller et al.\,(\cite{Hillier}), results in a slight decrease in wind 
emission at optical wavelengths. This decrease, however, was found to be less
than the increase of the wind emission due to a high $\beta$ value compared
to a wind with $\beta = 1.0$. We can, therefore, conclude that in clumped
winds with high $\beta$ values the effects discussed in this paper
are still present, but probably slightly weakened.

\section{Conclusions}

We investigated the influence of the thermal wind emission produced by 
free-free and free-bound processes to the total continuum of normal OB 
supergiants. While for winds with a velocity distribution following a 
$\beta$-law with $\beta$ typically in the range of 0.8 to 1.0, no influence of 
the wind at optical wavelengths is expected, the situation can be different 
when $\beta$ exceeds 1.0. High $\beta$ values reaching even 3.5 have recently
been found for many OB supergiants. Our investigations therefore concentrated 
on such high beta values and their influence on the wind continuum emission 
especially at optical wavelengths.

We found that the wind emission in the optical is generated within 
$2\,R_{*}$, only. This region is exactly the region where $\beta$ has its 
highest influence on the wind density structure. Since with increasing 
$\beta$ the wind is accelerated much more slowly than for $\beta = 1.0$, the 
density close to the stellar surface is strongly enhanced, leading to an 
enhanced production of especially free-bound emission at optical wavelengths.
At the same time the stellar emission, which passes through the wind on its way 
to the observer, is absorbed. These effects increase with increasing $\beta$.
The total continuum of OB supergiants, for which $\beta$ values higher than 1.0 
are found, can thus contain {\it non-negligible contributions of wind emission 
even at optical wavelengths}.

                                                                                
\begin{acknowledgements}
                                                                                
We thank the referee, Ian Howarth, for his comments.
M.K. acknowledges financial support from GA AV \v{C}R
under grant number KJB300030701.
                                                                                
\end{acknowledgements}
                                                                                

\end{document}